\documentclass[aps,pre,twocolumn,groupedaddress,amsmath,amssymb,floatfix]{revtex4}

\usepackage{float}
\usepackage{graphicx}
\usepackage{dcolumn}\usepackage{bm}
\usepackage{amssymb}

\newcommand{\sbentcore}{bent-core }

\newcommand{\Q}{\mathbf{Q}}
\newcommand{\xxi}{\hat{\boldsymbol{\xi}}}
\newcommand{\ee}{\hat{\boldsymbol{\eta}}}
\newcommand{\PP}{\mathbf{P}}
\newcommand{\rr}{\mathbf{r}}

\newcommand{\kk}{\mathbf{k}}

\newcommand{\e}{\mathbf{M}}

\providecommand{\orderparam}[3]{\overline{\Delta^{(#1)}_{#2,#3}}}
\providecommand{\dg}[1]{#1^{\circ}}

\begin{document}

\title{ Modulated nematic structures induced by  chirality and steric polarization}

\author{Lech Longa}
\email[e-mail address:]{lech.longa@uj.edu.pl}
\affiliation{Marian Smoluchowski Institute of Physics, Department
of Statistical Physics, Jagiellonian University, \L{}ojasiewicza 11,
Krak\'ow, Poland}
\author{Grzegorz Paj\c{a}k}
\email[e-mail address:]{grzegorz@th.if.uj.edu.pl}
\affiliation{Marian Smoluchowski Institute of Physics, Department
of Statistical Physics, Jagiellonian University, \L{}ojasiewicza 11,
Krak\'ow, Poland}

 \date{\today}

\begin{abstract}
What kind of one-dimensional modulated nematic structures (ODMNS)
can form nonchiral and chiral bent-core and dimeric materials?
Here, using Landau-deGennes theory of nematics, extended to account
for molecular steric polarization, we study a possibility of formation
of ODMNS, both in nonchiral and intrinsically chiral liquid crystalline
materials. Besides nematic and cholesteric phases, we find four bulk ODMNS
for nonchiral materials, two of which have not been reported so far.
These new structures are longitudinal ($N_{LP}$) and transverse ($N_{TP}$)
periodic waves where the polarization field being periodic in one dimension
stays parallel and perpendicular, respectively, to the wave vector. The other
two phases have all characteristic features of the twist-bend nematic
phase ($N_{TB}$) and the splay-bend nematic phase ($N_{SB}$), but their
fine structure appears more complex than that considered so far. The presence
of molecular chirality converts nonchiral $N_{TP}$ and  $N_{SB}$ into new $N_{TB}$ phases.
Interestingly, the nonchiral $N_{LP}$ phase can stay stable even in the presence of
intrinsic molecular chirality. Exemplary phase diagrams provide further insights into
the relative stability of these new modulated nematic structures.
 \end{abstract}

 \maketitle
Until very recently only four classes of nematics were recognized:
\emph{(i)} uniaxial and \emph{(ii)} biaxial nematics for nonchiral liquid crystalline
materials and \emph{(iii)} cholesteric and \emph{(iv)} blue phases for chiral liquid crystals
\cite{deGennesBook}. The most surprising recent discovery is the identification
of the fifth. In this new phase, found in liquid crystalline systems of
chemically \emph{achiral} dimers \cite{PanovNTB,CestariNTB,BorshchNTB,ChenBananyPNAS},
bent-core mesogens \cite{GleesonGoodby,ChenBananyPRE2014}
and their hybrids \cite{hybridbentcores}, the molecules
are arranged to form a helical superstructure with nanoscale periodicity.
This periodicity is about two orders of magnitude
shorter than typically found in cholesteric and blue phases
of ordinary chiral materials. As the molecular centers
of mass are distributed randomly in space the structure belongs
to the nematic class.
It coined the name twist-bend nematic phase ($N_{TB}$).
Contrary to cholesterics, the director in $N_{TB}$ is not perpendicular
to the helix axis, but
precesses on a cone, with the helical axis parallel to the cone's axis.
As formation of $N_{TB}$ does not require any presence of molecular
chirality experimentally coexisting domains of opposite chirality are observed.
So far the $N_{TB}$ phase is stabilized below uniaxial nematic phase
($N$) as a result of first order
$N-N_{TB}$ phase transition on the temperature scale. Hence, we observe
a fundamentally new phenomenon namely the \emph{spontaneous chiral
symmetry breaking} within the nematic class of materials.

Very recently, chiral asymmetric dimers were also studied  \cite{ZepGoreckaChiralSB,Gorecka2015}.
Here the intrinsic molecular chirality is an extra  factor giving
rise to overall a sequence of up to seven distinct nematic phases. As the
constituent molecules are intrinsically chiral the highest-temperature phase is
the cholesteric phase ($N^*$) or a blue phase. Stable phases observed at
lower temperatures  are variants of $N_{TB}$ with pitch which is larger
than one in achiral $N_{TB}$ \cite{Gorecka2015}.

The issue of stable $N_{TB}$
and $N_{SB}$ structures has  been addressed
at theoretical level in two important papers \cite{MeyerFlexopolarization,DozovNTB}.
With the aid of symmetry arguments, supplemented by second-order elasticity theory
of the director field,  Meyer \cite{MeyerFlexopolarization} and later on Dozov \cite{DozovNTB},
have analyzed some consequences of spontaneous local bend or splay deformations of the
nematic director on the polar organization of the molecules - the so called
flexoelectric effect.  While the cholesterics can fill the space with homogeneous
twist it appears that the corresponding bend state should always be associated
either with some twist or splay. According to this picture
 the uniform nematic phase would then become unstable
to the formation of a modulated phase, which could be either
$N_{TB}$ or $N_{SB}$ \cite{DozovNTB}. 
A prerequisite to such behaviour would be the sign change of the (effective)
bend Frank elastic constant, $K_3$. Indeed, $K_3$ determined experimentally
in $N$ \cite{BorshchNTB,AdlemLuckhurstNTB} is anomalously small as
the transition to $N_{TB}$ is approached.

Recently, Shamid  \emph{et al.} \cite{SelingerNTB2013}  have developed
Landau theory and lattice simulations of polar order and director bend deformations,
correlating flexoelectricity, negative $K_3$ and stability of $N_{TB}$
and $N_{SB}$ phases. They predicted a second-order phase transition from
high-temperature N phase to low-temperature $N_{TB}$ or $N_{SB}$.
At the transition, the effective $K_3$ changes sign and the
corresponding structure develops modulated polar order, averaging to zero
globally. All phases are assumed uniaxial and described entirely using director
and polarization fields.

The purpose of this Letter is to investigate how nematics can self-organize
into ODMNS
for nonchiral and intrinsically chiral V-shaped molecules, using first-principles,
symmetry-based generalization of
Landau-deGennes theory of nematics. We assume that the second-rank $3\times 3$
traceless and symmetric alignment tensor field, $ {\mathbf{Q}}({\mathbf{r}})$,
is the primary order parameter accounting for nematic order \cite{deGennesBook}.
It permits that locally a system is described by a tripod
of orthonormal directors $\{ \hat{\mathbf{n}}, \hat{\mathbf{l}},\hat{\mathbf{m}} \}$
and corresponding eigenvectors $\{ \lambda_n, \lambda_l, \lambda_m \}$.
Identifying the full biaxial field $\mathbf{Q}$ with primary order parameter
of nematics, rather than its $\hat{\mathbf{n}}$-part only \cite{DozovNTB},
should also clarify whether biaxiality ($\lambda_n \ne \lambda_l \ne \lambda_m$) is relevant for $N_{TB}$,
for we know that chiral nematic phases of at least intrinsically
chiral mesogens are all biaxial \cite{lechChiralIcosa}. Important theoretical
questions are thus about structure characterization
of ODMNS, both for non-chiral and intrinsically chiral materials
\cite{ZepGoreckaChiralSB,Gorecka2015}.

For the modeling of spontaneous chiral symmetry breaking as observed in $N_{TB}$  the
$\mathbf{Q}$-tensor alone is not sufficient.
In the lowest order scenario we need, in addition to $\mathbf{Q}$,
at least one secondary order parameter, which can be either a first-rank (polar) field,
$\mathbf{P}(\mathbf{r})$ \cite{lechChiralFlexo}, or a third-rank tensor field $\mathbf{T}(\mathbf{r})$,
invariant with respect to tetrahedral point group symmetry \cite{FelTetrahedratic,multitensor,%
BrandTetrahedratic,longaTetra2009,tetraDuality,longaTetra2012,longaTetra2013APP}.
The difference between these scenarios would be the polar order for $\mathbf{Q}$ and
$\mathbf{P}$ couplings \cite{lechChiralFlexo} and lack of polarity but the presence
of nonlinear dielectric tensor for $\mathbf{Q}$ and $\mathbf{T}$.
Here we focus on one-dimensional modulated structures as induced by
$\mathbf{Q}$ and $\mathbf{P}$. We look systematically
into extended Landau-deGennes-Ginzburg (LdeG) free energy expansion and discuss the
role played by various symmetry-allowed, lowest-order couplings. The essential
ordering mechanism towards one-dimensional modulated structures
will then be identified with the help of bifurcation and numerical analyses.

Before we proceed further it is important to realize
that the polar field, $\mathbf{P}$, does not need to be of
electrostatic origin. Bend-core molecules are primarily polar due to their "V" shape
while bimesogens acquire steric polarization in their conformational states. Such
\emph{steric dipoles} are present even in the absence of electrostatic dipoles.
In a densely
packed environment, we expect that these entropic, excluded volume interactions,
rather than charge moments, define the local order, such as $\mathbf{P}$.
Recently, Greco and Ferrarini have
studied systems composed of crescent-shaped molecules interacting through
a purely repulsive potential
providing strong support for entropy-driven
$N-N_{TB}$ transition \cite{2015prlGreco}.

We start by introducing the minimal coupling, LdeG free energy
per volume, constructed as a power-series expansion in
$\mathbf{Q}$ and $\mathbf{P}$, and their derivatives. It can be decomposed as
\begin{equation}\label{eq:freeEnergy}
  F= \sum_{i=2}^{4} F_i=\frac{1}{V}\sum_{i=2}^{4}\int_V\left( f_{iQ}+f_{iP}+f_{iQP}\right) \mathrm{d}^{3}{\rr},
\end{equation}
where  $f_{iX}$ are the free energy densities constructed out of
the order parameters $\{X\}$ and contributing to $F_i$ in
$i-$th order.
By taking suitable units of energy, length,  $\Q$ and  $\mathbf{P}$
and disregarding electric field,  magnetic field and  surface terms
the general form of $f_{iQ}$ up to fourth-order in $\Q$ can be written as
\begin{eqnarray}  \label{eq:ldegQ}
 \hspace{-0.1cm}f_{2Q} &=&
    \frac{1}{4}
       \left[ t_Q\, \mathrm{Tr}({\Q}^{2})+ (\mathbf{\nabla}\otimes \Q)\cdot
       (\mathbf{\nabla}\otimes \Q)  \right.\nonumber\\
             & & \left. + \rho (\mathbf{\nabla}\cdot \Q)\cdot
             (\mathbf{\nabla}\cdot \Q) - 2 \kappa \, \mathbf{Q \cdot (\nabla\times \Q)} \right] \\
 f_{3Q}&+&f_{4Q}=
    - \sqrt{6}\, B\, \mathrm{Tr}({\Q}^{3})
    + \mathrm{Tr}({\Q}^{2})^{2},
\end{eqnarray}
where, as usual, $t_Q$ is the reduced temperature associated with $\Q$; $\kappa$ is the
reduced chirality,
proportional to the wave vector of the cholesteric phase and  $\rho$ is the
relative elastic constant.
Likewise, the polar parts $f_{iP}$ are
\begin{eqnarray}  \label{ldegP}
 f_{2P} &+& f_{3P} + f_{4P} =
    \frac{1}{4}
       \left[ t_P \, \mathbf{P}^{2} +
              (\mathbf{\nabla \otimes P})\cdot(\mathbf{\nabla \otimes P})
              \right. \nonumber \\ &+& \left. a_c (\mathbf{\nabla \cdot P})^2 -2 \kappa_P \, \mathbf{P \cdot (\nabla \times P)}
       \right] + a_4 (\mathbf{P}^{2})^2
\end{eqnarray}
Clearly, for \emph{electric dipoles}  $\mathbf{\nabla \times P}$ would vanishes in (\ref{ldegP},\ref{ldegQP})
while the $a_c$-term should be replaced by direct interactions between charge distributions.
However, for purely steric dipoles, associated with excluded volume
interactions \cite{2015prlGreco} these terms are all present.
In particular, for intrinsically chiral materials that develop \emph{steric}
polar order the chiral parameters
$\kappa$,  $\kappa_P$ and $\kappa_{QP}$ are all nonzero.

General, orientational properties of polar biaxial liquid crystals,
characterized by $\mathbf{Q}$ and $\mathbf{P}$, were analyzed in
\cite{lechElastic1,lechChiralFlexo}. In particular the flexopolarization
couplings were identified and classified.
Assuming deformations to appear only in a quadratic part
of the free energy and neglecting surface terms
the lowest-order cross-coupling terms $f_{iQP}$ are
\begin{eqnarray}  \label{ldegQP}
 f_{2QP} &=&   -\frac{1}{4}\left[
 e_P \mathbf{\mathbf{P}\cdot( \nabla \cdot Q})  + 2\kappa_{QP} (\mathbf{\nabla \cdot Q})\cdot
 (\mathbf{\nabla \times P}) \right] \hspace{0.6cm} \\
 f_{3QP} &=& - \lambda P_\alpha Q_{\alpha\beta} P_\beta \\ \label{ldegQP4}
 f_{4QP}&=&  \lambda_1 P_\alpha Q_{\alpha\beta}^2 P_\beta + \lambda_2 \mathbf{P}^2 Tr(\mathbf{Q}^2).
\end{eqnarray}
The LdeG expansion (\ref{eq:ldegQ}-\ref{ldegQP4}) is the minimal coupling theory for systems
described in terms of $\PP$ and $\Q$, where $\PP$ is of steric origin.
Our objective is to identify possible ODMNS that minimize
$F([{\Q (\mathbf{r} ),\mathbf{P(r)}}])$
for arbitrary $t_Q$ and $t_P>0$, and for fixed values of the material parameters.
By taking $t_P>0$ we assume from the start that $\mathbf{P}$ is secondary order parameter.
A brief account of what to expect from such theory has already been presented long ago
in \cite{lechChiralFlexo}, where we indicated on a possibility of
flexopolarization-induced periodic, one-dimensional structures.
A more quantitative analysis
of modulated nematic structures that can be driven by (flexo-)polarization
is found in \cite{AlexanderFlexoBP,SelingerPolarBluePhases2014,MishaNTB&NSB}.
In their theory  Alexander and Yeomans
\cite{AlexanderFlexoBP} showed that applying an electric field
to a sample with a large flexoelectric response can stabilize $N_{SB}$
and a flexoelectric blue phase.
Shamid, Allender and Selinger \cite{SelingerPolarBluePhases2014}, on the other hand,
have taken a simpler version of the expansion \cite{lechChiralFlexo} and showed that the
system can stabilize a polar analogue of chiral blue phases if  polar
order is allowed to be induced spontaneously.

In this letter we study inhomogeneous  nematics with inhomogeneity
propagating in one spatial direction, both for nonchiral and intrinsically chiral materials.
We show that modulated phases,
identified so far as a result of polar coupling, do not exhaust all
possibilities that the theory (\ref{eq:freeEnergy}-\ref{ldegQP4}) allows for.
Our ultimate goal will be to look  into fine structure of these phases and
clarify the role played by biaxiality.
In order to address these issues we explore the bifurcation theory
supplemented by numerical minimization and
identify global minima of $F$,
Eq.~(\ref{eq:freeEnergy}), within the ODMNS family,
leaving a more complex issue of stable blue phases to our future studies.

The above programme is realized in practice by
expanding $\Q(\rr )$ and $\PP(\rr )$ into plane waves of definite helicity:
%
%
 $       \Q(\rr) =
         \sum_{\kk}
               \sum^{2}_{m=-2}
                  {Q_{m}(\kk) }
                  \exp(\mathrm{i}\: {\kk\cdot\rr} )\,
                  \e^{[2]}_{m,{\hat{\kk} }}
$,
%
%
 $                \PP(\rr) =
         \sum_{\kk}
               \sum^{1}_{m=-1}
                  {P_{m}(\kk) }
                  \exp(\mathrm{i}\: {\kk\cdot\rr}
                       )\,
                  \e^{[1]}_{m,{\hat{\kk} }}.
$
$\,$  Here ${\kk}$ are wave-vectors, $P_{m}({\kk})$ and
$Q_{m}({\kk})$
are the variational parameters in the
free energy expansion, and
 $\e^{[1]}_{\pm 1,{\hat{\kk} }} =\mp\frac{1}{\sqrt{2}} ( \xxi \pm \mathrm{i} \ee)$,
  $\e^{[1]}_{0,{\hat{\kk} }} =\hat{\kk}$,
  $\e^{[2]}_{\pm 2,{\hat{\kk} }} = \frac{1}{2}( \xxi \pm \mathrm{i} \ee)\otimes ( \xxi \pm \mathrm{i} \ee)$,
  $\e^{[2]}_{\pm 1,{\hat{\kk} }} = \mp\frac{1}{2}\left[ ( \xxi \pm \mathrm{i} \ee)
  \otimes\hat{\kk} +  \hat{\kk} \otimes ( \xxi \pm \mathrm{i} \ee)\right]$, and finally
  $\e^{[2]}_{0,{\hat{\kk} }} = \frac{1}{\sqrt{6}}( 3 \hat{\kk}\otimes \hat{\kk} - \mathbf{1})$
%
%
%
%
are the spin $L = 1, 2$ spherical tensors represented in an orthonormal,
right handed local coordinate system $\{ \xxi,\ee, \hat{\kk} \}$ with ${\hat{\kk}}$
as quantization axis. Two amplitudes of opposite helicity out of
$\{ Q_{m}(\kk)$, $P_{m}(\kk) \}$ can be taken real  due to invariance
of $\PP$ and $\Q$ with respect to uniform translations in 3D and  global
rotations about $\kk$. We choose $\mathfrak{Im}Q_{\pm 1}=0$.
The reality condition: $\{\Q(\rr) = \Q(\rr)^{*},\PP(\rr)=\PP(\rr)^*\}$ additionally implies that
$Q_m(-\kk)=(-1)^{m} Q_m(\kk)^{*}$ and $P_m(-\kk)=(-1)^{m+1} P_m(\kk)^{*}$.
%
The selection of ${\kk}$ and  relevant amplitudes $\{ Q_{m}(\kk)$, $P_{m}(\kk) \}$
is fixed by the bifurcation analysis  \cite{longajcp,longaBif} while their numerical values are found
by the subsequent minimization of $F$.

In the vicinity of the isotropic phase the dominant contributions to $F$, Eq.~(\ref{eq:freeEnergy}), comes from
$F_{2}$ which, for the ODMNS structures, is
\begin{eqnarray} \label{F2}
  F_2 &=& \sum_{n}\sum_{m=-2}^{2}\left\{ A_m(|n|k) |Q_m(n)|^2
  + (1-\delta_{m^2,4}) \left[ \right.\right. \nonumber \\&& \hspace{-0cm} B_m(|n|k) |P_m(n)|^2
+
    \frac{\mathrm{i}}{2}C_m(|n|k) \left(
    Q_m(n)\, P_m^*(n) \right.
   \nonumber \\&&
\left. \left.\left. \hspace{3cm} - Q_m^*(n) \, P_m(n) \right) \,\right]\right\},
\end{eqnarray}
where
%
\[  A_m(|n|k) = \left[ {t_Q} +  n^2 k^2 +\frac{\rho (4-m^2)}{6} n^2 k^2 - \kappa m |n| k \right]/4   \]

\begin{eqnarray}
\hspace{-1.1cm}  B_m(|n|k) &=& \left[ {t_P} + n^2 k^2 + a_c (1-m^2) n^2 k^2 \right. \nonumber \\
 &&\hspace{3.5cm} \left.
  - 2 \kappa_P m |n| k   \right]/4
\end{eqnarray}
\[ C_m(|n|k) =  -\frac{1}{4}\left(e_P \sqrt{\frac{4-m^2}{6}} |n| k +  \sqrt{2}\kappa_{QP} m n^2 k^2 \right).\]
Here $\kk $ is replaced by $ n k, \, n=0,\pm 1,...$, $P_m(\kk)$ by $P_m(n)$ and
$Q_m(\kk)$ by $Q_m(n)$; $\delta_{i,j}$ is the Kronecker delta.
Setting $\partial F/\partial Q_m(n) = \partial F/\partial P_m(n) = \partial F/\partial k = 0$ determines
the equilibrium value of the amplitudes and $k$-vector  for given material parameters.
Since explicit dependence on $k$ appears only in second-order contributions,
we have
%
%
$  \partial F/\partial k = \partial F_2/\partial k = 0$.
%
%
Note that the $F_2$-terms linear in $k$ promote modulated structures,
among which are ODMNS. As it turns out an interesting class of ODMNS
can already be identified  by studying  a simpler model
where $\kappa_P=\kappa_{QP}= \lambda_1=\lambda_2=0$. In what follows we shall consider
this simpler case leaving analysis of the full model, along with the issue of stable blue phases, to
our forthcoming publications. In addition we take $t_P>0$ and $\kappa\ge 0$.
For thermodynamic stability it is also necessary
that $\rho > -\frac{3}{2}$ and  $1+a_c>0$. Additionally, $a_4$ must be positive if $\lambda \ne 0$.

We shall now proceed by analyzing the case of $\lambda=a_4=0$
and later discuss the effect of $\lambda\ne 0$.
For the first-mentioned case the polarization field appears only in $F_2$ and,
hence, the condition $\partial F/\partial P_m(n) = 0$ can be solved
for $P_m(n)$ given fixed $Q_m(n)$. It yields
\begin{equation}\label{eqPm}
  P_m(n)=-\mathrm{i} \frac{C(m,|n|k)}{2 B(m,|n|k)}\, Q_m(n), \,\,\, m=0,\pm1.
\end{equation}
Substituting (\ref{eqPm}) back to $F$ we obtain the effective free energy that
still has to be minimized with respect to $Q_m(n)$.
Only $F_2$ is modified by this substitution. It reads
\begin{eqnarray}\label{F2elP}
  F_{2,eff}=\sum_{n} \sum_{m=-2}^2\left[
  A(m,|n|k)-\frac{C(m,|n|k)^2 }{4 B(m,|n|k)}(1-\delta_{m^2,4})
  \right] \nonumber \\
 && \hspace{-7cm}\times |Q_m(n)|^2 = \frac{1}{V}\int_V \mathrm{f}_{eff}(\Q,\partial \Q) \mathrm{d}^{3}{\rr}.
\end{eqnarray}
Note that the leading elastic part of $\mathrm{f}_{eff}$ can again
be cast in form (\ref{eq:ldegQ}), but with $\rho$ being replaced
by $\rho-\frac{e_P^2}{4 t_P}$. Since $(\mathbf{\nabla}\cdot \Q)^2$
vanishes for twist deformations \cite{lechElastic1} the
flexopolarization must induce splay-bend instability for
$\rho-\frac{e_P^2}{4 t_P} \le -\frac{3}{2}$.
However, the $e_P$ term alone cannot bring about spontaneous chiral symmetry breaking.
For that we need sufficiently large $|\lambda |$.

\begin{figure}[ht]
    \begin{picture}(50,200)
    \put(-75, 50){
      \includegraphics{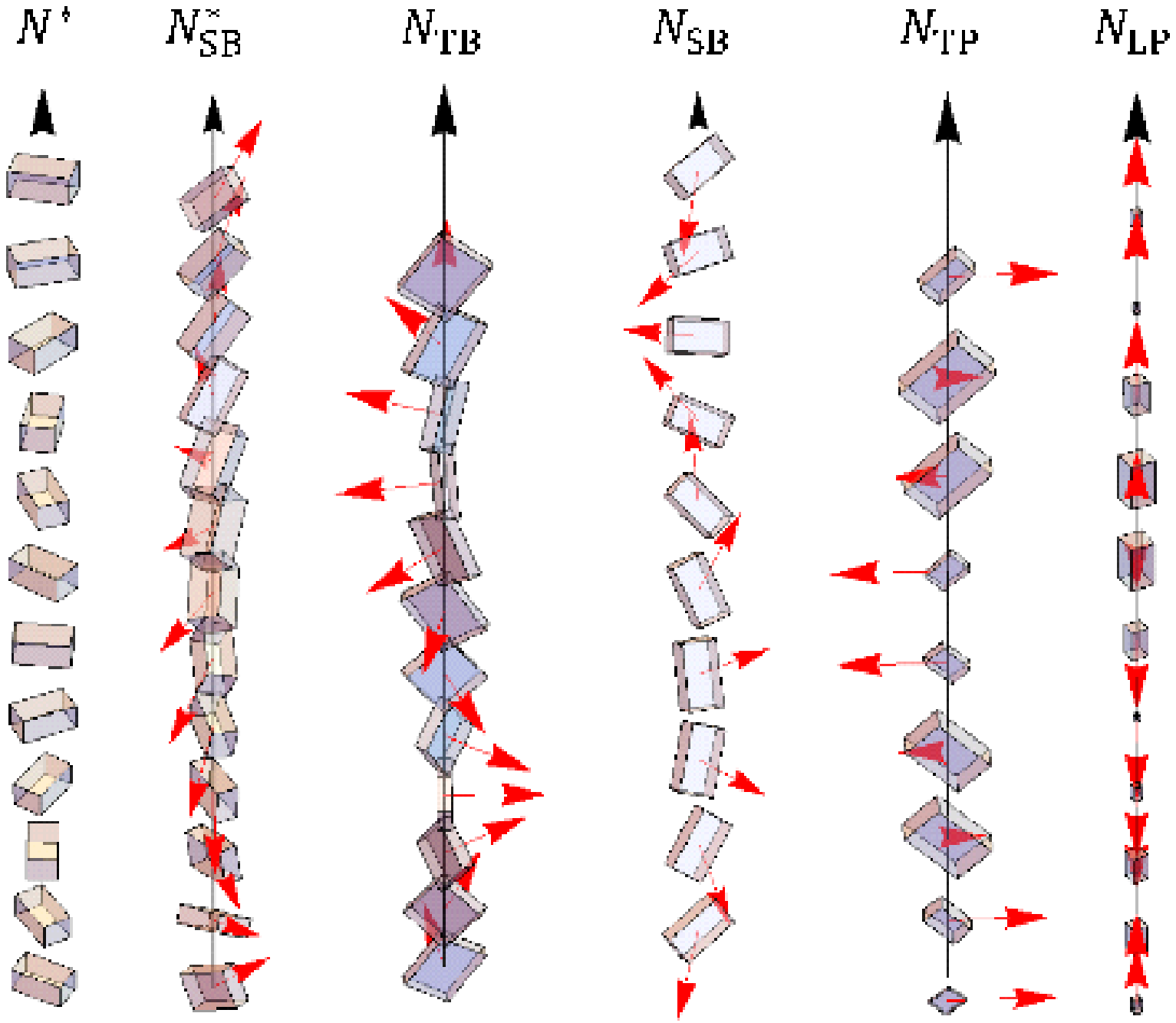}
    }
    \end{picture}
\caption[]{(Color online) ODMNS predicted by the theory
(leading nonzero amplitudes are given in braces):
$N_{LP}$ $\{Q_0(1)$, $P_0(1)$, $\mathfrak{Re}Q_0(0)\}$,
$N_{TP}$ $\{\mathfrak{Re} Q_{\pm1}(1)$, $\mathfrak{Re}Q_0(0)$, $\mathfrak{Re}Q_2(0)$, $\mathfrak{Im}P_{\pm 1}(1)\}$,
$N_{TB}$ $\{$as  in $N_{TP}\}$,
$N_{SB}$ $\{$as in $N_{TP}$, $\mathfrak{Im}Q_{\pm2}(1)$, $\mathfrak{Im}Q_0(1)\}$,
$N^*_{SB}$ $\{$as in $N_{SB}\}$ and
$N^*$ $\{\mathfrak{Im}Q_{\pm2}(1)$, $\mathfrak{Re}Q_0(0)\}$.
Lengths of cuboid edges are proportional to the moduli of eigenvalues of $\Q$.
Red arrows represent $\PP$ and black arrow is the direction of $\kk$. }
\label{structures}
\end{figure}

We shall now seek for ODMNS that can be stabilized as result of
a phase transition from the isotropic phase. The general method
is to analyse  the nonlinear equations $\partial F/\partial Q_m(n) = \partial F/\partial P_m(n) =0$
for the amplitudes $Q_m(n)$ and $P_m(n)$ using the bifurcation ana\-ly\-sis
(BA) \cite{longajcp,longaBif}. We apply this method to identify the leading amplitudes
in the expansion of $\mathbf{Q}$ and $\mathbf{P}$ close to a phase transition
from the isotropic phase ($I$), where $Q_m(n)=P_m(n)=0$, to an ordered ODMNS.
The procedure is straightforward for  $\lambda=a_4=0$. In the zeroth-order of BA
the amplitudes $Q_m(n)$ are close to their isotropic values and,
hence, governed by the $F_{2,eff}$ part of $F$. Since $F_{2,eff}$ is in its diagonal form we
can identify five different ODMNS, each characterized by the single $Q_m(1)$ mode
of helicity $m$, that bifurcate from $I$ at temperatures $t_Q=t_m$,
where $t_m$ equals to  $t_Q$
at which the coefficient in front of $|Q_m(1)|^2$ in Eq.~(\ref{F2elP})
vanishes. The corresponding wavectors, $k_m$, are determined from
$\partial F_2/\partial k = 0$  with $P_m(n)$
given by Eq.~(\ref{eqPm}).
The maximal of the temperatures $t_m$ represents a potential transition temperature from
$I$ to ODMNS for a continuous phase transition and spinodal for a first order phase transition.
Explicitly, the $m=\pm 2$ modes bifurcate from $I$ when
%
   $t_Q=t_{\pm 2}=\kappa ^2$ and $k_{\pm 2}=\pm \kappa$. 
%
The condensation of $m=\pm 1 $ modes occurs when $t_{\pm 1}$ and  $k_{\pm 1}$
satisfy the implicit relations
%
$  t_{\pm 1}=
  \frac{k_{\pm 1} \left(e_P^2 k_{\pm 1} - 4 \left(k_{\pm 1}^2+t_P\right)
  \left(\mp 2 \kappa +(\rho +2) k_{\pm 1}\right) \right)}{8 \left(k_{\pm 1}^2+t_P\right)}
$
%
and
$
  \kappa =\frac{k_{\pm 1} \left(\mp e_P^2 t_P \pm 4 (\rho +2)
  \left(k_{\pm 1}^2+t_P\right)^2\right)}{4 \left(k_{\pm 1}^2+t_P\right)^2}.
$
%
They can be resolved for non-chiral materials ($\kappa=0$) giving
%
 $ t_{\pm 1}=\frac{1}{8} \left(e_P^2-4 \sqrt{(\rho +2) e_P^2 t_P}+4 (\rho +2) t_P\right)$
%
and
%
$  k_{\pm 1}=\sqrt{\sqrt{\frac{e_P^2 t_P}{4( \rho +2)}}-t_P} $,
which are satisfied for
$ 0<t_P<\frac{e_P^2}{4 ( \rho +2)}$.
%
Finally,
 $ t_0=\frac{e_P^2-2 \sqrt{2 (2 \rho +3) e_P^2 t_P}+2 (2 \rho +3) t_P}{6 \left(a_c+1\right)}$
%
and
%
$  k_0=\sqrt{\frac{ \sqrt{\frac{2 e_P^2 t_P}{2 \rho +3}}-2 t_P}{2 (a_c+1)}}$, for $0<t_P<\frac{e_P^2}{2 (2 \rho +3)}$.
%
%
We should mention that the bifurcation from $I$ to $N$ takes place at  $t_Q=0$.
First-order BA, consistent with modelling of the $N^*$ phase \cite{lechChiralIcosa},
allows to identify the next to leading amplitudes $Q_m(n),\, n=0,1$
of ODMNS that couple to those given above through $F_3$ and $F_4$. A subsequent minimization
of $F$ either with respect to so identified trial states or with respect to all amplitudes of $|n|\le 1$ gives
(consistently) six different ODMNS, shown in  Fig.~\ref{structures}.

\begin{figure}
 \includegraphics[width=60mm]{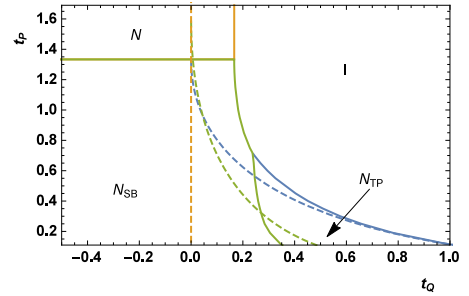}%
 \caption{
  \label{diagram1}
 (Color online) Phase diagram for $\rho=1$, $\kappa=0$, $e_P=-4$,
 $B={1}/{\sqrt{6}}$ and $a_c=2$. Solid lines are obtained from
 numerical minimization while dashed curves are bifurcation
 temperatures from the isotropic phase.
 }
\end{figure}

\begin{figure}
 \includegraphics[width=60mm]{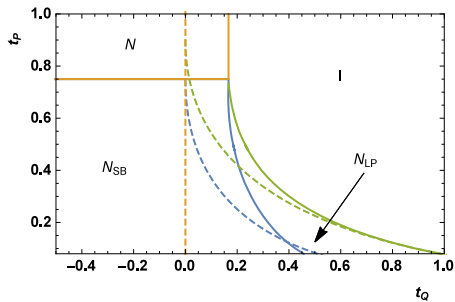}%
 \caption{
  \label{diagram2}
 (Color online) Phase diagram for $\rho=1$, $\kappa=0$, $e_P=-3$, $B=\frac{1}{\sqrt{6}}$
 and $a_c=-\frac{1}{4}$. For further details see caption to Fig.~\ref{diagram1}.
 }
\end{figure}

\begin{figure}
 \includegraphics[width=60mm]{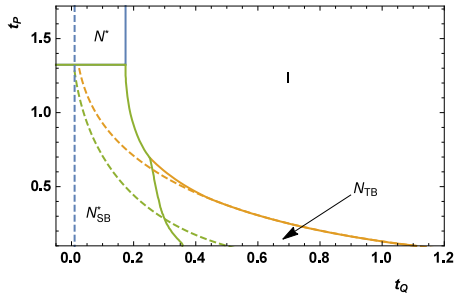}%
 \caption{
  \label{diagram3}
 (Color online) Phase diagram for $\rho=1$, $\kappa=\frac{1}{10}$,
 $e_P=-4$, $B=\frac{1}{\sqrt{6}}$ and $a_c=2$. Here all modulated phases
 are chiral. For further details see caption to Fig.~\ref{diagram1}.
 }
\end{figure}

\begin{figure}
 \includegraphics[width=60mm]{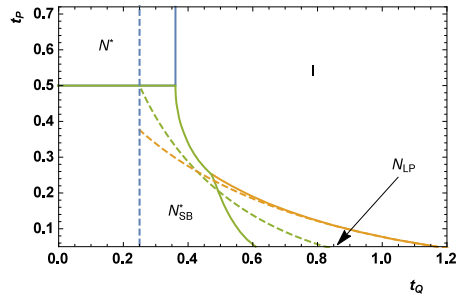}%
 \caption{
  \label{diagram4}
 (Color online) Phase diagram for $\rho=1$, $\kappa=\frac{1}{2}$,
 $e_P=-3$, $B=\frac{1}{\sqrt{6}}$ and $a_c=-\frac{1}{4}$.
 Despite nonzero intrinsic chirality there is a stable region of
 achiral modulated $N_{LP}$ structure. For further details see caption to Fig.~\ref{diagram1}.
 }
\end{figure}

Our  numerical minimization is carried out for the exemplary
sets of material parameters. The corresponding phase diagrams are shown in
Figs~(\ref{diagram1}-\ref{diagram4}). The bifurcation
temperatures from the isotropic phase
are also plotted as dashed lines.
All phase transitions involved are at least weakly first order although with
increasing $t_Q$ and decreasing $t_P$ the difference
between the bifurcation- and transition temperatures becomes
numerically negligible.

Figs~(\ref{diagram1},\ref{diagram2}) show new ODMNS structures
predicted by the model for $\kappa=0$.
These flexopolarization-induced nonchiral structures
are referred to as $N_{TP}$ and $N_{LP}$.
In $N_{TP}$ the polarization vector, Fig.~(\ref{structures}),
is always perpendicular to $\kk$ and $\hat{\mathbf{m}}$, and
$\mathfrak{Re} Q_{+1}(1)=-\mathfrak{Re} Q_{-1}(1)$.
The homogeneous nematic background ($\mathfrak{Re}Q_0(0)\ne 0$, $\mathfrak{Re}Q_2(0)\ne 0$) makes $N_{TP}$
locally biaxial with biaxiality modulated along $\kk$.
The $N_{LP}$ phase is constructed out of $m=0$ modes. Here
$\hat{\mathbf{n}}$ and $\PP$ are always parallel to $\kk$.
The phase is uniaxial and periodically changes between prolate and oblate.
In order to obtain the transition from $I$ to
$N_{LP}$ the $a_c-$ term, (Eq.\ref{F2}),
should be small positive or negative.
In the $N_{SB}$ phase an inhomogeneous  biaxial modulation is also generated.
Here $\PP$ is periodically modulated in the $\{\hat{\mathbf{n}}$,$\hat{\mathbf{l}}\}$-plane of splay-bend deformations.
The phase diagrams presented in Figs~(\ref{diagram3}-\ref{diagram4})
show changes induced by intrinsic molecular chirality for $\lambda=0$. Clearly, $\kappa\ne0$ results in replacing  $N$
by $N^{*}$, where $\PP=0$ and $\hat{\mathbf{n}}$ rotates about $\kk$ and remains everywhere perpendicular to $\kk$.
Changes also concern $N_{TP}$ and $N_{SB}$. $N_{TP}$ transforms into  $N_{TB}$,
Fig.~\ref{structures}, which is biaxial with $\hat{\mathbf{n}}$
precessing on the cone about $\kk$
and $\PP$ parallel to $\hat{\mathbf{m}}$ and perpendicular to $\kk$.
The $N_{SB}^{*}$ phase is also biaxial,  where two out of the three directors
posses twist deformations similar to the ones modelled in \cite{MishaNTB&NSB}.
$\PP$ in this phase is a linear combination of all three directors.
Interestingly, the nonchiral $N_{LP}$ phase can become stable even in intrinsically
chiral materials, Fig.~(\ref{diagram4}).

The $N_{TB}$ phase can be stabilized not only for  $\kappa\ne 0$, but primarily
when  $\lambda\ne0$ $(a_4>0)$. In order to obtain the $N_{TB}$ phase for
nonchiral materials the sign of $\lambda$ must be consistent
with the sign of $e_P$ and $|\lambda|$ should exceed
a threshold value. For example, if we take parameters of Fig.~\ref{diagram1} and $a_4=1$, the $N_{TB}$ phase
becomes stable for $\lambda \lesssim -0.4$. Calculations carried out for $\lambda=-1/2$ show
rich sequence of phase transitions: $I\leftrightarrow(N_{TP},N_{SB},N)\leftrightarrow N_{TB}$,
where phases in parentheses are optional. Although the $N_{TB}$ phases obtained for \emph{(a)} $\{\kappa\ne0, \lambda=0\}$ and  \emph{(b)} $\{\kappa=0, \lambda\ne0\}$
have the same symmetry, in the first case the structure of helicity $m=Sign(\kappa)$ minimizes $F$, while for the case \emph{(b)}
structures of opposite helicities $m=\pm1$ are of the same free energy, which is know as \emph{ambidextrous chirality}.
The cases \emph{(a)} and \emph{(b)} differ quantitatively,  \emph{i.e.} in periodicities and
biaxiality parameter \cite{lechChiralIcosa}.

In conclusion, LdeG phenomenological theory of nematics
extended to account for molecular steric polarization, stabilizes four bulk ODMNS
for nonchiral materials, two of which: $N_{LP}$ and $N_{TP}$ have not been reported so far.
The presence
of molecular chirality converts $N_{TP}$ into $N_{TB}$ and  $N_{SB}$ into new chiral $N^*_{SB}$,
but the nonchiral $N_{LP}$ phase can stay stable even in the presence of
intrinsic molecular chirality.

\begin{acknowledgements}
The authors thank Misha Osipov for stimulating discussions.
This work was supported by the Grant No. DEC-2013/11/B/ST3/04247 of the National Science Centre in Poland.
\end{acknowledgements}


\end{document}